\documentclass[11pt]{article}
\usepackage{amssymb,amsmath,amsfonts}
\usepackage{graphicx}
\usepackage{graphics}
\usepackage{eepic,epsfig}

=-1.3truecm \voffset = -2truecm \evensidemargin = 1.20cm
\begin{document}

\title{Electrostatic self-interaction in the spacetime of a global monopole
with finite core}
\author{E. R. Bezerra de Mello$^{1}$\thanks{%
E-mail: emello@fisica.ufpb.br}\, and A. A. Saharian$^{1,2}$\thanks{%
E-mail: saharian@ictp.it} \\
\\
\textit{$^1$Departamento de F\'{\i}sica-CCEN, Universidade Federal da Para%
\'{\i}ba}\\
\textit{58.059-970, Caixa Postal 5.008, Jo\~{a}o Pessoa, PB, Brazil}\vspace{%
0.3cm}\\
\textit{$^2$Department of Physics, Yerevan State University,}\\
\textit{375025 Yerevan, Armenia}}
\maketitle

\begin{abstract}
In this paper we calculate the induced electrostatic self-energy and
self-force for an electrically charged particle placed at rest in the
spacetime of a global monopole admitting a general spherically symmetric
inner structure to it. In order to develop this analysis we calculate the
three-dimensional Green function associated with this physical system. We
explicitly show that for points outside the monopole's core the self-energy
presents two distinct contributions. The first is induced by the non-trivial
topology of the global monopole considered as a point-like object. The
second is a correction induced by the non-vanishing inner structure
attributed to it. As an illustration of the general procedure the flower-pot
model for the region inside the monopole is considered. In this application
it is also possible to find the electrostatic self-energy for points in the
region inside the monopole. In the geometry of the global monopole with the
positive solid angle deficit, we show that for the flower-pot model the
electrostatic self-force is repulsive with respect to the core surface for
both exterior and interior regions.
\end{abstract}

\bigskip

{PACS number(s): 98.80.Cq, 14.80.Hv}

\newpage

\section{Introduction}

It is well known that different types of topological objects may have been
formed by the vacuum phase transition in the early Universe after Planck
time \cite{Kibble,V-S}. These include domain walls, cosmic strings and
monopoles. Global monopoles are heavy topological objects formed in the
phase transition of a system composed by a self-coupling iso-triplet scalar
field $\Phi ^{a}$ whose original global $O(3)$ symmetry is spontaneously
broken to $U(1)$. The scalar matter field plays the role of an order
parameter which outside the monopole's core acquires a non-vanishing value.
The global monopole was first introduced by Sokolov and Starobinsky \cite%
{Soko} and the gravitational effects of the global monopole have been
analyzed by Barriola and Vilenkin \cite{BV}. It has been shown that for
points far away from the monopole's center the corresponding geometry can be
described by the line element
\begin{equation}
ds^{2}=dt^{2}-dr^{2}-\alpha ^{2}r^{2}(d\theta ^{2}+\sin ^{2}\theta d\varphi
^{2})\ ,  \label{gm1}
\end{equation}%
with the parameter $\alpha ^{2}=1-8\pi G\eta ^{2}$ determined by the energy
scale $\eta $ where the global symmetry is spontaneously broken. It is of
interest to note that the effective metric produced in superfluid $^{3}%
\mathrm{He-A}$ by a monopole is described by line element (\ref{gm1}) with
the negative angle deficit, $\alpha >1$, which corresponds to the negative
mass of the topological object \cite{Volo98}.

Many of treatments in the investigation of physical effects around a global
monopole deal mainly with the case of the idealized point-like monopole
geometry described by line element (\ref{gm1}) for all values of the radial
coordinate. However, the realistic global monopole has a characteristic core
radius determined by the symmetry braking scale at which the monopole is
formed. The calculation of the metric tensor in the region inside the global
monopole would require the knowledge of the behavior of the energy-momentum
tensor associated with the scalar field $\Phi ^{a}$, which on the other hand
requires the knowledge of the components of the metric tensor, providing, in
this way, a non solvable integral equation \cite{Mello1}. In this paper we
shall not go into the details of this calculation. Instead, we shall
consider a simplified model described by two sets of the metric tensor for
two distinct regions, continuous at a spherical shell of radius $a$. In the
exterior region corresponding to $r>a$, the line element is given by (\ref%
{gm1}), while in the interior region, $r<a$, the geometry is described by
the static spherically symmetric line element%
\begin{equation}
ds^{2}=u^{2}(r)dt^{2}-v^{2}(r)dr^{2}-w^{2}(r)(d\theta ^{2}+\sin ^{2}\theta
d\varphi ^{2})\ .  \label{gm2}
\end{equation}%
At the boundary of the core the functions $u(r)$, $v(r)$, $w(r)$ satisfy the
conditions
\begin{equation}
u(a)=v(a)=1,\ w(a)=\alpha a\ .  \label{bound}
\end{equation}%
By introducing a new radial coordinate $\tilde{r}=w(r)$ with the core center
at $\tilde{r}=0$, the angular part of the line element (\ref{gm2}) is
written in the standard Minkowskian form. With this coordinate, in general,
we will obtain non-standard angular part in the exterior line element.

Many years ago, Linet \cite{Linet} and Smith \cite{Smith}, independently,
have shown that an electrically charged particle placed at rest in the
spacetime of an idealized cosmic string becomes subjected to a repulsive
self-interaction. This self-interaction is a consequence of the distortion
of the particle's fields caused by the planar angle deficit associated with
the conical geometry. Also it was shown in \cite{Mello2} that a linear
electric or magnetic sources in the spacetime of a cosmic string parallel to
the latter, become subject to induced self-interactions. More recently the
problem of the induced electrostatic self-energy in the spacetime of a thick
cosmic string has been considered in \cite{NV}. Analogously what happens in
the cosmic string spacetime, a point-like electrically charged particle
placed at rest in the spacetime of an idealized global monopole, becomes
also subjected to a repulsive self-interaction \cite{Mello3}. In the present
paper we shall continue in this line of investigation. We shall consider the
induced electrostatic self-energy and self-force associated with a
point-like charged particle placed at rest in the spacetime of a global
monopole with a finite core described by line element (\ref{gm2}). The
corresponding results specify the conditions under which we can ignore the
details of the interior structure and approximate the effect of the global
monopole by the idealized model. The analysis of quantum vacuum effects for
a scalar field in the model under consideration has been developed in \cite%
{Mello4}. It was shown that these effects are composed by the sum of a
point-like monopole and core-induced parts. Moreover, adopting a specific
model for the monopole's core, the flower-pot one, explicit calculations for
the vacuum polarization effects in the exterior and interior regions have
been done.

This paper is organized as follows. Writing the Maxwell equations in the
spherically symmetric spacetime, in section \ref{sec:Outside} we calculate
the three-dimensional Green function for points outside and inside the
monopole's core. As a consequence, we provide a general expression for the
electrostatic self-energy and the related self-force. We shall see that in
the exterior region the corresponding expressions are composed by two parts.
The first ones are induced by the global monopole considered as a point-like
object, while the second parts are induced by the non-vanishing inner
structure attributed to it. As an illustration of the general results
obtained, in section \ref{sec:Flower}\ we consider the flower-pot model for
the region inside the core. In this model, we explicitly calculate the
self-energy in exterior and interior regions and describe its behavior in
various asymptotic regions of the parameters. In section $5$ we present our
conclusions and more relevant remarks.

\section{Self-energy outside the monopole core}

\label{sec:Outside}

The main objective of this paper is to evaluate the electrostatic
self-energy and the self-force for a point-like charged particle at rest,
induced by the spacetime geometry associated with a global monopole with the
core of finite radius \footnote{%
For the electrostatic self-force on a charged test particle held stationary
outside a Schwarzschild black hole see~\cite{Vile79}.}.We will assume that
in the region inside the monopole core the geometry is described by line
element (\ref{gm2}), and in the exterior region we have the standard line
element (\ref{gm1}) with the solid angle deficit $4\pi (1-\alpha ^{2})$. For
the covariant components of the electromagnetic four-vector potential, $%
A_{i} $, from the Maxwell equations we have
\begin{equation}
\partial _{k}\left[ \sqrt{-g}g^{im}g^{kn}\left( \partial _{m}A_{n}-\partial
_{n}A_{m}\right) \right] =-4\pi \sqrt{-g}j^{i},  \label{Meq1}
\end{equation}%
where $j^{i}$ is the four-vector electric current density. For a point-like
particle at rest with coordinates $\mathbf{r}_{0}=(r_{0},\theta _{0},\varphi
_{0})$, in the coordinate system corresponding to the line element (\ref{gm2}%
), the static four-vector current and potential read: $j^{i}=(j^{0},0,0,0)$
and $A_{n}=(A_{0},0,0,0)$. The only nontrivial component of (\ref{Meq1}) is
the $i=0$ one with
\begin{equation}
j^{0}(x)=q\frac{\delta (\mathbf{r}-\mathbf{r}_{0})}{\sqrt{-g}}\ ,  \label{J0}
\end{equation}%
where $q$ is the charge of the particle. So, in the spherically symmetric
spacetime defined by (\ref{gm2}), the differential equation obeyed by $A_{0}$
reads:%
\begin{equation}
\partial _{r}\left( \frac{w^{2}}{uv}\partial _{r}A_{0}\right) -\frac{v}{u}%
\widehat{\mathbf{L}}^{2}A_{0}=-\frac{4\pi q}{\sin \theta }\delta (\mathbf{r}-%
\mathbf{r}_{0}),  \label{Meq2}
\end{equation}%
with $\widehat{\mathbf{L}}$ being the operator of the angular momentum. The
solution of this equation can be written in terms of the Green function
associated with the differential operator defined by the left-hand side, as
follows:
\begin{equation}
A_{0}(\mathbf{r})=4\pi qG(\mathbf{r},\mathbf{r}_{0})\ ,  \label{A0}
\end{equation}%
with the equation for the Green function%
\begin{equation}
\left[ \partial _{r}\left( \frac{w^{2}}{uv}\partial _{r}\right) -\frac{v}{u}%
\widehat{\mathbf{L}}^{2}\right] G(\mathbf{r},\mathbf{r}_{0})=-\frac{\delta
(r-r_{0})}{\sin \theta }\delta (\theta -\theta _{0})\delta (\varphi -\varphi
_{0}).  \label{Green-a}
\end{equation}

Having the electrostatic self-potential for the charge we can evaluate the
corresponding self-force by using the standard formula%
\begin{equation}
f_{\mathrm{el}}^{i}(\mathbf{r}_{0})=qg^{ik}F_{km}u^{m}=q\frac{g^{ik}}{u}%
\partial _{k}A_{0}|_{\mathbf{r}=\mathbf{r}_{0}}=4\pi q^{2}\frac{g^{ik}}{u}%
\lim_{\mathbf{r}\rightarrow \mathbf{r}_{0}}\left[ \partial _{k}G(\mathbf{r},%
\mathbf{r}_{0})\right] .  \label{feli}
\end{equation}%
An alternative way to obtain the self-force is to consider first the
electrostatic self-energy given by \cite{Linet,Smith}
\begin{equation}
U_{\mathrm{el}}(\mathbf{r}_{0})=qA_{0}(\mathbf{r}_{0})/2=2\pi q^{2}\lim_{%
\mathbf{r}\rightarrow \mathbf{r}_{0}}G(\mathbf{r},\mathbf{r}_{0})\ ,
\label{SE}
\end{equation}%
and then to derive the force on the base of the formula%
\begin{equation}
f_{\mathrm{el}}^{i}(\mathbf{r}_{0})=\frac{g^{ik}}{u}\partial _{k}U_{\mathrm{%
el}}(\mathbf{r}_{0}).  \label{feli2}
\end{equation}%
In accordance to the Synge's theorem, formulae (\ref{feli}) and (\ref{feli2}%
) lead to the same result for the self-force.

In formulae (\ref{feli}) and (\ref{SE}) the limit is divergent. To obtain a
finite and well defined result for the self-force, we should apply some
renormalization procedure for the Green function. The procedure that we
shall adopt is the standard one (see, for instance, \cite{Birrell}): we
subtract from the Green function the terms in the corresponding
DeWitt-Schwinger adiabatic expansion which are divergent in the coincidence
limit. So, we define the renormalized Green function as
\begin{equation}
G_{\mathrm{ren}}(\mathbf{r},\mathbf{r}_{0})=G(\mathbf{r},\mathbf{r}_{0})-G_{%
\mathrm{DS}}^{\mathrm{(div)}}(\mathbf{r},\mathbf{r}_{0})\ .  \label{Grenn}
\end{equation}%
In this way the renormalized self-energy, $U_{\mathrm{el,ren}}(\mathbf{r}%
_{0})$, and self-force, $f_{\mathrm{el,ren}}^{i}(\mathbf{r}_{0})$, are
obtained by the formulae (\ref{feli}) and (\ref{SE}) with the replacement $G(%
\mathbf{r},\mathbf{r}_{0})\rightarrow G_{\mathrm{ren}}(\mathbf{r},\mathbf{r}%
_{0})$. Note that here the subtraction of the divergent part of the Green
function corresponds to the renormalization of the particle mass.

Taking into account the spherical symmetry of the problem, we may present
the Green function as the expansion
\begin{equation}
G(\mathbf{r},\mathbf{r}_{0})=\sum_{l=0}^{\infty
}\sum_{m=-l}^{l}g_{l}(r,r_{0})Y_{l}^{m}(\theta ,\varphi )Y_{l}^{m\ast
}(\theta _{0},\varphi _{0})\ ,  \label{Green-b}
\end{equation}%
with $Y_{l}^{m}(\theta ,\varphi )$ being the ordinary spherical harmonics.
Substituting (\ref{Green-b}) into (\ref{Green-a}) and using the well known
closure relation for the spherical harmonics, we arrive at the differential
equation for the radial function:
\begin{equation}
\left[ \frac{d}{dr}\left( \frac{w^{2}}{uv}\frac{d}{dr}\right) -\frac{v}{u}%
l(l+1)\right] g_{l}(r,r_{0})=-\delta (r-r_{0}).  \label{g}
\end{equation}%
As the functions $u(r)$, $v(r)$, $w(r)$ are continuous at $r=a$, from (\ref%
{g}) it follows that the function $g_{l}(r,r_{0})$ and its first radial
derivative are also continuous at this point. The function $g_{l}(r,r_{0})$
is continuous for $r=r_{0}$ as well. The junction of the first radial
derivative at $r=r_{0}$ is obtained by the integration of (\ref{g}) about
this point:%
\begin{equation}
\frac{dg_{l}(r,r_{0})}{dr}|_{r=r_{0}+}-\frac{dg_{l}(r,r_{0})}{dr}%
|_{r=r_{0}-}=-\frac{u(r_{0})v(r_{0})}{w^{2}(r_{0})}.  \label{derjunc}
\end{equation}

In the region inside the core, we denote by $R_{1l}(r)$ and $R_{2l}(r)$ the
linearly independent solutions of the homogeneous equation corresponding to (%
\ref{g}). We shall assume that the function $R_{1l}(r)$ is regular at the
core center $r=r_{c}$ and that the solutions are normalized by the Wronskian
relation
\begin{equation}
R_{1l}(r)R_{2l}^{\prime }(r)-R_{1l}^{\prime }(r)R_{2l}(r)=-\frac{u(r)v(r)}{%
w^{2}(r)}.  \label{Wronin}
\end{equation}%
In the region outside the core the linearly independent solutions to the
corresponding homogeneous equation are the functions $r^{\lambda _{1}}$ and $%
r^{\lambda _{2}}$, where
\begin{equation}
\lambda _{1,2}=-\frac{1}{2}\pm \frac{1}{2\alpha }\sqrt{\alpha ^{2}+4l(l+1)}\
\ .  \label{lam12}
\end{equation}%
Now, we can write $g_{l}(r,r_{0})$ as a function of the radial coordinate $r$
in the separate regions $[r_{c},\min (r_{0},a))$, $(\min (r_{0},a),\max
(r_{0},a))$, and $(\max (r_{0},a),\infty )$ as a linear combination of the
above mentioned solutions with arbitrary coefficients. The requirement of
the regularity at the core center and at the infinity reduces the number of
these coefficients to four. They are determined by the continuity condition
at the monopole's core boundary and by the matching conditions at $r=r_{0}$.
In this way we find the following expressions%
\begin{eqnarray}
g_{l}(r,r_{0}) &=&\frac{(ar_{0})^{\lambda _{1}}R_{1l}(r)}{\alpha ^{2}\left[
aR_{1l}^{\prime }(a)-\lambda _{2}R_{1l}(a)\right] },\;\mathrm{for}%
\;r\leqslant a,  \label{gl1in} \\
g_{l}(r,r_{0}) &=&\frac{r_{<}^{\lambda _{1}}r_{>}^{\lambda _{2}}}{\alpha
^{2}(\lambda _{1}-\lambda _{2})}\left[ 1-\left( \frac{a}{r_{<}}\right)
^{\lambda _{1}-\lambda _{2}}D_{1l}(a)\right] ,\;\mathrm{for}\;r\geqslant a,
\label{gl1out}
\end{eqnarray}%
in the case $r_{0}>a$, and
\begin{eqnarray}
g_{l}(r,r_{0})
&=&R_{1l}(r_{<})R_{2l}(r_{>})-R_{1l}(r_{0})R_{1l}(r)D_{2l}(a),\;\mathrm{for}%
\;r\leqslant a,  \label{gl2in} \\
g_{l}(r,r_{0}) &=&\frac{a^{\lambda _{1}}r^{\lambda _{2}}R_{1l}(r_{0})}{%
\alpha ^{2}\left[ aR_{1l}^{\prime }(a)-\lambda _{2}R_{1l}(a)\right] },\;%
\mathrm{for}\;r\geqslant a,  \label{gl2out}
\end{eqnarray}%
in the case $r_{0}<a$. In these formulae, $r_{<}=\min (r,r_{0})$ and $%
r_{>}=\max (r,r_{0})$, and we have used the notation%
\begin{equation}
D_{jl}(a)=\frac{aR_{jl}^{\prime }(a)-\lambda _{j}R_{jl}(a)}{aR_{1l}^{\prime
}(a)-\lambda _{2}R_{1l}(a)},\;j=1,2.  \label{Djl}
\end{equation}

First let us consider the case when the charge is situated outside the
monopole's core ($r_{0}>a$). Substituting the function (\ref{gl1out}) into (%
\ref{Green-b}), we see that the Green function is presented in the form of
the sum%
\begin{equation}
G(\mathbf{r},\mathbf{r}_{0})=G_{\mathrm{m}}(\mathbf{r},\mathbf{r}_{0})+G_{%
\mathrm{c}}(\mathbf{r},\mathbf{r}_{0}),  \label{Ggumar}
\end{equation}%
where%
\begin{equation}
G_{\mathrm{m}}(\mathbf{r},\mathbf{r}_{0})=\frac{1}{4\pi \alpha r_{>}}%
\sum_{l=0}^{\infty }\frac{2l+1}{\sqrt{\alpha ^{2}+4l(l+1)}}\frac{%
r_{<}^{\lambda _{1}}}{r_{>}^{\lambda _{1}}}P_{l}(\cos \gamma ),  \label{Gm}
\end{equation}%
is the Green function for the geometry of a point-like global monopole, and
the term
\begin{equation}
G_{\mathrm{c}}(\mathbf{r},\mathbf{r}_{0})=-\frac{1}{4\pi \alpha }%
\sum_{l=0}^{\infty }\frac{(2l+1)D_{1l}(a)}{\sqrt{\alpha ^{2}+4l(l+1)}}%
(rr_{0})^{\lambda _{2}}P_{l}(\cos \gamma )\ \ ,  \label{Gc}
\end{equation}%
is induced by non-trivial structure of the core. In formulae (\ref{Gm}) and (%
\ref{Gc}), the $\gamma $ is the angle between the directions $(\theta
,\varphi )$ and $(\theta _{0},\varphi _{0})$, and $P_{l}(x)$ represents the
Legendre polynomials. It can be seen that (\ref{Gm}) coincides, up to the
redefinition of the radial variable $r\rightarrow \alpha r$, with the
expression found in \cite{Mello3}, for the case of a point-like global
monopole spacetime. The part (\ref{Gc}) depends on the structure of the core
through the radial function $R_{1l}(r)$.

As we have already mentioned, the induced self-energy is obtained from the
renormalized Green function taking the coincidence limit. We can observe
that for points with $r>a$, the core-induced term (\ref{Gc}) is finite in
the coincidence limit and the divergence appears in the point-like monopole
part only. So, in order to provide a well defined finite value to (\ref{SE}%
), we have to renormalize Green function $G_{\mathrm{m}}(\mathbf{r},\mathbf{r%
}_{0})$ only:
\begin{equation}
G_{\mathrm{ren}}(r_{0},r_{0})=G_{\mathrm{m,ren}}(r_{0},r_{0})+G_{\mathrm{c}%
}(r_{0},r_{0})\ .  \label{Gren}
\end{equation}%
As explained before, to find $G_{\mathrm{m,ren}}(r_{0},r_{0})$, we subtract
from (\ref{Gm}) the terms in the corresponding DeWitt-Schwinger adiabatic
expansion which are divergent in the coincidence limit:%
\begin{equation}
G_{\mathrm{m,ren}}(r_{0},r_{0})=\lim_{\mathbf{r}\rightarrow \mathbf{r}_{0}}%
\left[ G_{\mathrm{m}}(\mathbf{r},\mathbf{r}_{0})-G_{\mathrm{m,DS}}^{\mathrm{%
(div)}}(\mathbf{r},\mathbf{r}_{0})\right] .  \label{Gmren}
\end{equation}%
The part $G_{\mathrm{m,DS}}^{\mathrm{(div)}}(\mathbf{r},\mathbf{r}_{0})$ is
found from the general formula given, for instance, in \cite{Birrell},
specifying the parameters for the problem under consideration. It can be
seen that here the first term in the DeWitt-Schwinger expansion contributes
only to the divergent part. For simplicity, taking the separation of the
points along the radial direction only ($\gamma =0$), we find
\begin{equation}
G_{\mathrm{m,DS}}^{\mathrm{(div)}}(r,r_{0})=\frac{1}{4\pi |r-r_{0}|}\ .
\label{Gmdiv}
\end{equation}%
Now, by using formulae (\ref{Gm}) and (\ref{Gmdiv}), one obtains%
\begin{equation}
G_{\mathrm{m,ren}}(r_{0},r_{0})=\frac{1}{4\pi r_{0}}\lim_{t\rightarrow 1}%
\left[ \frac{1}{\alpha }\sum_{l=0}^{\infty }\frac{2l+1}{\sqrt{\alpha
^{2}+4l(l+1)}}\ t^{\lambda _{1}}-\frac{1}{1-t}\right] ,  \label{Gmren1}
\end{equation}%
where $t=r_{<}/r_{>}$. To evaluate the limit on the right, we note that%
\begin{equation}
\lim_{t\rightarrow 1}\left( \frac{1}{\alpha }\sum_{l=0}^{\infty }t^{l/\alpha
+1/2\alpha -1/2}-\frac{1}{1-t}\right) =0.  \label{RelLim}
\end{equation}%
On the basis of this relation, replacing in (\ref{Gmren1}) $1/(1-t)$ by the
first term in the brackets in (\ref{RelLim}), we find%
\begin{equation}
G_{\mathrm{m,ren}}(r_{0},r_{0})=\frac{S(\alpha )}{4\pi r_{0}},
\label{Gmren2}
\end{equation}%
where we have introduced the notation%
\begin{equation}
S(\alpha )=\frac{1}{\alpha }\sum_{l=0}^{\infty }\left[ \frac{2l+1}{\sqrt{%
\alpha ^{2}+4l(l+1)}}-1\right] .  \label{falfa}
\end{equation}%
The function $S(\alpha )$ is positive (negative) for $\alpha <1$ ($\alpha >1$%
) and, hence, the corresponding self-force is repulsive (attractive).
Developing a series expansion in the parameter $\eta ^{2}=1-\alpha ^{2}$, we
can see that
\begin{equation*}
\alpha S(\alpha )=\sum_{n=1}^{\infty }\frac{(\pi \eta )^{2n}}{2(n!)^{2}}%
|B_{2n}|(1-2^{-2n}),
\end{equation*}%
where $B_{n}$ are the Bernoulli numbers. The leading term in the expression
on the right is $\pi (1-\alpha ^{2})/16$. For large values $\alpha $ the
main contribution into the series in (\ref{falfa}) comes from large values $l
$. Replacing the summation by the integration we can see that in the limit $%
\alpha \rightarrow \infty $ the function $S(\alpha )$ tends to the limiting
value $-1/2$. For small values $\alpha $, $\alpha \ll 1$, the main
contribution comes from the term $l=0$ and one has $S(\alpha )\approx
1/\alpha ^{2}$.

Combining formulae (\ref{SE}), (\ref{Gc}), and (\ref{Gmren2}), for the
renormalized electrostatic self-energy we get
\begin{equation}
U_{\mathrm{el,ren}}(r_{0})=\frac{q^{2}S(\alpha )}{2r_{0}}-\frac{q^{2}}{%
2\alpha r_{0}}\sum_{l=0}^{\infty }\frac{(2l+1)D_{1l}(a)}{\sqrt{\alpha
^{2}+4l(l+1)}}\left( \frac{a}{r_{0}}\right) ^{\sqrt{1+4l(l+1)/\alpha ^{2}}}\
.  \label{U-out}
\end{equation}%
The second term of the renormalized self-energy provides a convergent series
for $r_{0}>a$. Here the dependence of the self-energy on the core structure
appears through the function $D_{1l}(a)$. For large distances from the core,
$r_{0}\gg a$, the main contribution into the core-induced part comes from
the term $l=0$ and one has%
\begin{equation}
U_{\mathrm{el,ren}}(r_{0})\approx \frac{q^{2}}{2r_{0}}\left[ S(\alpha )-%
\frac{aD_{10}(a)}{\alpha ^{2}r_{0}}\right] .  \label{Ulargedist}
\end{equation}%
The self-force is obtained from (\ref{U-out}) by using formula (\ref{feli2}):%
\begin{equation}
\mathbf{f}_{\mathrm{el,ren}}(\mathbf{r}_{0})=U_{\mathrm{el,ren}}(r_{0})\frac{%
\mathbf{r}_{0}}{r_{0}^{2}}-\frac{q^{2}\mathbf{r}_{0}}{2\alpha ^{2}r_{0}^{3}}%
\sum_{l=0}^{\infty }(2l+1)D_{1l}(a)\left( \frac{a}{r_{0}}\right) ^{\sqrt{%
1+4l(l+1)/\alpha ^{2}}}.  \label{felout}
\end{equation}%
In accordance with the symmetry of the problem, the self-force has only a
radial component. We can see that the same result for the self-force is
obtained on the basis of formula (\ref{feli}).

Now we turn to the case when the charge is inside the core, $r_{0}<a$. The
corresponding Green function is obtained from (\ref{gl2in}) and is written
in the form%
\begin{equation}
G(\mathbf{r},\mathbf{r}_{0})=G_{0}(\mathbf{r},\mathbf{r}_{0})+G_{\alpha }(%
\mathbf{r},\mathbf{r}_{0}),  \label{Gin}
\end{equation}%
where
\begin{equation}
G_{0}(\mathbf{r},\mathbf{r}_{0})=\frac{1}{4\pi }\sum_{l=0}^{\infty
}(2l+1)R_{1l}(r_{<})R_{2l}(r_{>})P_{l}(\cos \gamma )\ ,  \label{G0in}
\end{equation}%
is the Green function for the background geometry described by the line
element (\ref{gm2}) for all values $r_{c}\leqslant r<\infty $, and the term%
\begin{equation}
G_{\alpha }(\mathbf{r},\mathbf{r}_{0})=-\frac{1}{4\pi }\sum_{l=0}^{\infty
}(2l+1)R_{1l}(r_{0})R_{1l}(r)D_{2l}(a)P_{l}(\cos \gamma ),  \label{Galfain}
\end{equation}%
is due to the global monopole geometry in the region $r>a$. For the points
away from the core boundary the latter is finite in the coincidence limit.
The self-energy for the charge inside the core is written in the form
\begin{equation}
U_{\mathrm{el,ren}}(r_{0})=2\pi q^{2}G_{0,\mathrm{ren}}(r_{0},r_{0})-\frac{%
q^{2}}{2}\sum_{l=0}^{\infty }(2l+1)D_{2l}(a)R_{1l}^{2}(r_{0}),
\label{Urenin}
\end{equation}%
where%
\begin{equation}
G_{0,\mathrm{ren}}(r_{0},r_{0})=\lim_{\mathbf{r}\rightarrow \mathbf{r}_{0}}%
\left[ G_{0}(\mathbf{r},\mathbf{r}_{0})-G_{0,\mathrm{DS}}^{\mathrm{(div)}}(%
\mathbf{r},\mathbf{r}_{0})\right] .  \label{G0renin}
\end{equation}%
The only contribution in the divergent part of the Green function comes from
the first term of the DeWitt-Schwinger expansion. Note that near the center
of the core one has $R_{1l}(r_{0})\propto (r_{0}-r_{c})^{l}$ and the main
contribution into the second term on the right of (\ref{Urenin}) comes from
the term with $l=0$. Substituting the self-energy given by (\ref{Urenin})
into formula (\ref{feli2}), we obtain the self-force for the charge inside
the monopole core.

\section{Flower-pot model}

\label{sec:Flower}

As we have mentioned in the Introduction, it is not possible to provide a
closed expression to the metric tensor in the region inside the global
monopole. A few years ago, Harari and Lousto \cite{HL} proposed a simplified
model for the monopole where the region inside the core is described by the
de Sitter geometry. The vacuum polarization effects associated with a
massless scalar field in the region outside the core of this model have been
investigated in \cite{Mello5}. As to the cosmic string spacetime, in the
literature two different models have been adopted to describe the geometry
inside core: the ballpoint-pen model proposed independently by Gott and
Hiscock \cite{Gott}, which corresponds to replacing the conical singularity
at the string axis by a constant curvature spacetime in the interior region,
and flower-pot model \cite{BA}, where the curvature is concentrated on a
ring and the spacetime inside the string is flat. Adopting the latter model
for a global monopole, in \cite{Mello4} we were able to provide exact
expressions for the vacuum polarization effects associated with a massive
scalar field in both exterior and interior regions of the monopole
spacetime. So, motivated by this result we decided, as an illustration of
the general results described above, to consider this model in the present
analysis of the induced electrostatic self-interaction. In line element (\ref%
{gm2}), taking $u(r)=v(r)=1$, from the zero curvature condition one finds $%
w(r)=r+\mathrm{const}$. The value of the constant here is found from the
continuity condition for the function $w(r)$ at the core boundary which
gives $\mathrm{const}=(\alpha -1)a$. Hence, in the flower-pot model the
interior line element has the form%
\begin{equation}
ds^{2}=dt^{2}-dr^{2}-\left[ r+(\alpha -1)a\right] ^{2}(d^{2}\theta +\sin
^{2}\theta d^{2}\varphi )\ .  \label{intflow}
\end{equation}%
In terms of the radial coordinate $r$ the origin is located at $%
r=r_{c}=(1-\alpha )a$. Defining $\tilde{r}=r+(\alpha -1)a$, the line element
takes the standard Minkowskian form. From the Israel matching conditions for
the metric tensors corresponding to (\ref{gm1}) and (\ref{intflow}), we find
the nonzero components of the corresponding surface energy-momentum tensor
located on the bounding surface $r=a$ \cite{Mello4}:%
\begin{equation}
\tau _{0}^{0}=2\tau _{2}^{2}=2\tau _{3}^{3}=\frac{1/\alpha -1}{4\pi Ga}.
\label{surfemtflow}
\end{equation}%
Note that the surface energy density is positive for $\alpha <1$.

Now in the flower-pot model we can express the renormalized Green function
in the region outside the monopole core by taking into account that in the
interior region we have the linearly independent solutions%
\begin{equation}
R_{1l}(r)=\tilde{r}^{l},\;R_{2l}(r)=\tilde{r}^{-l-1}/(2l+1).
\label{R12Flower}
\end{equation}%
So, from formula (\ref{U-out}), the self-energy in the exterior region reads%
\begin{eqnarray}
U_{\mathrm{el,ren}}(r_{0}) &=&\frac{q^{2}S(\alpha )}{2r_{0}}+\frac{%
2q^{2}(1-\alpha )}{\alpha r_{0}}\sum_{l=0}^{\infty }\frac{l(2l+1)}{\sqrt{%
\alpha ^{2}+4l(l+1)}}  \notag \\
&&\times \frac{(a/r_{0})^{\sqrt{1+4l(l+1)/\alpha ^{2}}}}{\left[ \sqrt{\alpha
^{2}+4l(l+1)}+\alpha +2l\right] ^{2}}\ .  \label{UrenFlow}
\end{eqnarray}%
The second term on the right of this formula is positive for $\alpha <1$ and
negative for $\alpha >1$. Combining this with the properties of the function
$S(\alpha )$ discussed in the previous section, we conclude that the
electrostatic self-energy is positive for $\alpha <1$ and negative for $%
\alpha >1$. The corresponding self-force is directly found from (\ref{felout}%
) and is repulsive in the first case and attractive in the second one. For
large distances from the monopole core the main contribution into the
core-induced part comes from the $l=1$ term (note that the $l=0$ term
vanishes) and we have%
\begin{equation}
U_{\mathrm{el,ren}}(r_{0})\approx \frac{q^{2}}{2r_{0}}\left[ S(\alpha )+%
\frac{12(1-\alpha )}{\alpha \sqrt{\alpha ^{2}+8}}\frac{(a/r_{0})^{\sqrt{%
1+8/\alpha ^{2}}}}{\left( \sqrt{\alpha ^{2}+8}+\alpha +2\right) ^{2}}\right]
,\;a\ll r_{0}.  \label{UrenFlowerlarge}
\end{equation}%
In this limit the effects induced due to the finite core are
relatively suppressed by the factor $(a/r_{0})^{\sqrt{1+8/\alpha
^{2}}}$. The core-induced part in (\ref{UrenFlow}) diverges at the
core boundary, $r_{0}=a$. The surface divergences in the
coincidence limit of the Green function in field theories on
background of manifolds with boundaries are well investigated in
the literature related to the Casimir effect. Noting that for
points near the boundary the main contribution into
(\ref{UrenFlow}) comes from large values $l$, to the leading order
we find
\begin{equation}
U_{\mathrm{el,ren}}(r_{0})\approx q^{2}\frac{\alpha -1}{8\alpha a}\ln \left[
1-\left( a/r_{0}\right) ^{1/\alpha }\right] ,\;  \label{GrenFlow}
\end{equation}%
and the self-energy is dominated by the core-induced part.

\begin{figure}[tbph]
\begin{center}
\epsfig{figure=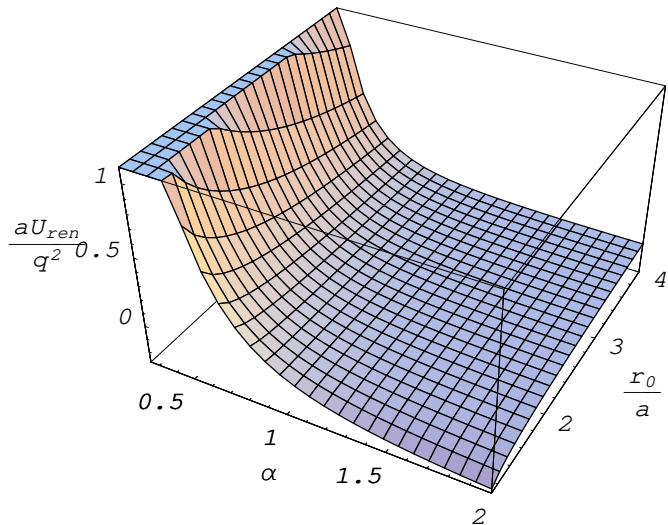,width=9.cm,height=8.cm}
\end{center}
\caption{Electrostatic self-energy in the flower-pot model for a charge
outside the monopole core as a function of the monopole parameter $\protect%
\alpha $ and rescaled radial coordinate $r_{0}/a$.}
\label{fig1}
\end{figure}

Now we turn to the investigation of the core-induced part in asymptotic
regions of the parameter $\alpha $. For large values $\alpha $ we replace
the summation over $l$ by the integration and to the leading order we find%
\begin{equation}
U_{\mathrm{el,ren}}(r_{0})\approx -\frac{q^{2}}{4r_{0}}\left[
1+\int_{1}^{\infty }dx\,\frac{\sqrt{x^{2}-1}-x+1}{\sqrt{x^{2}-1}+x+1}\left(
\frac{a}{r_{0}}\right) ^{x}\right] ,\;\alpha \gg 1.  \label{Ualflarge}
\end{equation}%
Hence, in the limit $\alpha \rightarrow \infty $ the renormalized self-force
tends to the finite limiting value. For small values $\alpha $, the main
contribution into the core-induced part of the self-energy comes from the
mode $l=1$ and this part is exponentially suppressed by the factor $\exp [2%
\sqrt{2}\ln (a/r_{0})/\alpha ]/\alpha $. We recall that in this limit the
point-like monopole part behaves like $1/\alpha ^{2}$ and, hence, it
strongly dominates. In figure \ref{fig1} we have plotted the electrostatic
self-energy in the flower-pot model for a charge outside the core versus the
parameter $\alpha $ and the radial coordinate of the charge.

Now we turn to the investigation of the self-energy in the flower-pot model
for the particle inside the monopole core. Substituting the functions (\ref%
{R12Flower}) into formulae (\ref{G0in}) and (\ref{Galfain}), for the
corresponding Green functions in the interior region one finds%
\begin{eqnarray}
G_{0}(\mathbf{r},\mathbf{r}_{0}) &=&\frac{1}{4\pi |\mathbf{r}-\mathbf{r}_{0}|%
},\;  \label{G0Flower} \\
G_{\alpha }(\mathbf{r},\mathbf{r}_{0}) &=&\frac{1}{4\pi \alpha a}%
\sum_{l=0}^{\infty }\frac{2l+2-\alpha -\sqrt{\alpha ^{2}+4l(l+1)}}{2l+\alpha
+\sqrt{\alpha ^{2}+4l(l+1)}}\frac{(\tilde{r}{_{0}}\tilde{r})^{l}}{(\alpha
a)^{2l}}P_{l}(\cos \gamma ),  \label{GalfaFlower}
\end{eqnarray}%
Because in the flower-pot model the geometry in the region inside the
monopole is a Minkowski one, we have $G_{0,\mathrm{DS}}^{\mathrm{(div)}}(%
\mathbf{r},\mathbf{r}_{0})=G_{0}(\mathbf{r},\mathbf{r}_{0})$ and, hence, $%
G_{0,\mathrm{ren}}(r_{0},r_{0})=0$. Finally, the electrostatic self-energy
in the region inside the monopole core reads:
\begin{equation}
U_{\mathrm{el,ren}}(r_{0})=2\pi q^{2}G_{\mathrm{ren}}(r_{0},r_{0})=\frac{%
q^{2}}{2\alpha a}\sum_{l=0}^{\infty }\frac{2l+2-\alpha -\sqrt{\alpha
^{2}+4l(l+1)}}{2l+\alpha +\sqrt{\alpha ^{2}+4l(l+1)}}\left( \frac{\tilde{r}{%
_{0}}}{\alpha a}\right) ^{2l}.  \label{Gr-in1}
\end{equation}%
As in the case of the exterior region, this self-energy is positive for $%
\alpha <1$ and negative for $\alpha >1$. The corresponding self-force is
easily found from relation (\ref{feli2}) and is repulsive with respect to
the boundary of the monopole core in the first case and attractive in the
second case. Near the core center the main contribution into the self-energy
comes from the lowest modes and one has%
\begin{equation}
U_{\mathrm{el,ren}}(r_{0})\approx \frac{q^{2}}{2\alpha a}\left[ \frac{%
1-\alpha }{\alpha }+\frac{4-\alpha -\sqrt{\alpha ^{2}+8}}{2+\alpha +\sqrt{%
\alpha ^{2}+8}}\left( \frac{\tilde{r}{_{0}}}{\alpha a}\right) ^{2}\right] .
\label{Unearcent}
\end{equation}

\begin{figure}[tbph]
\begin{center}
\epsfig{figure=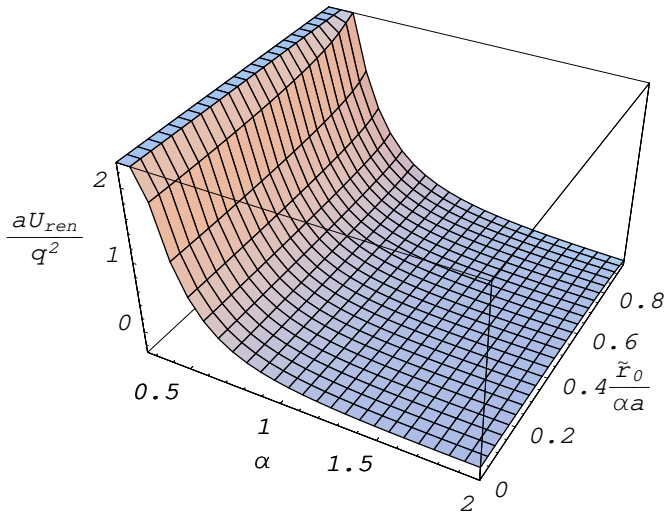,width=9.cm,height=8.cm}
\end{center}
\caption{Electrostatic self-energy in the flower-pot model for a charge
inside the monopole core as a function of the monopole parameter $\protect%
\alpha $ and rescaled radial coordinate $\tilde{r}_{0}/\protect\alpha a$.}
\label{fig2}
\end{figure}

As in the exterior case, on the core surface the self-energy given by (\ref%
{Gr-in1}) diverges. Under the condition $1/\ln (\alpha a/\tilde{r}{_{0}})\gg
\alpha $, the leading term in the corresponding asymptotic expansion is
given by the formula
\begin{equation}
U_{\mathrm{el,ren}}(r_{0})\approx q^{2}\frac{\alpha -1}{8\alpha a}\ln \left(
1-\frac{\tilde{r}_{0}}{\alpha a}\right) \ .  \label{GrenFlow1}
\end{equation}%
For large values $\alpha $, assuming that the ratio $\tilde{r}_{0}/\alpha a$
is fixed and $\alpha \gg 1/\ln (\alpha a/\tilde{r}{_{0}})$ (note that $%
\alpha a$ is the core radius for an internal Minkowskian observer), from (%
\ref{Gr-in1}) we find%
\begin{equation}
U_{\mathrm{el,ren}}(r_{0})\approx -\frac{q^{2}}{2\alpha a}\frac{1}{1-(\tilde{%
r}_{0}/\alpha a)^{2}}.  \label{Uinlargealf}
\end{equation}%
For small values $\alpha $ with the fixed value of the ratio $\tilde{r}%
_{0}/\alpha a$, the leading term in the self-energy is obtained substituting
$\alpha =0$ in the fraction of the expression under the summation sign in (%
\ref{Gr-in1}). In figure \ref{fig2} we have presented the dependence of the
electrostatic self-energy in the flower-pot model for a charge inside the
core as a function on $\alpha $ and $\tilde{r}_{0}/\alpha a$.

Up to now we have considered the electrostatic self-interaction for a
point-like particle. Similar results with the replacement of the electric
charge by the magnetic one can be obtained for a point-like magnetic charge
in the global monopole spacetime. In particular, from the results given
above it follows that with the regard for finite core, the composite system
of global and magnetic monopoles proposed in \cite{Beze02} can be stable.
Note that in the model with a point-like global monopole the corresponding
system can have a problem with the stability \cite{Achu03}. Assuming that
the gravitational field of the particle can be adequately described by the
Newtonian potential in the global monopole spacetime, we can also evaluate
the gravitational self-interaction in the flower-pot model by using the same
Green function. The corresponding self-energy is related to the
electrostatic self-energy by the formula%
\begin{equation}
U_{\mathrm{gr,ren}}(r_{0})=-(GM^{2}/q^{2})U_{\mathrm{el,ren}}(r_{0}),
\label{GravEn}
\end{equation}%
where $G$ is the Newton gravitational constant and $M$ is the mass of the
particle.

\section{Concluding remarks}

The objective of this paper was to analyze the induced self-energy
and the self-force for a point-like electric charge placed at rest
in the spacetime of a global monopole considering a non-trivial
inner structure for the core. As it was previously shown
\cite{Mello3}, for a idealized core of a point-like global
monopole, the above quantities present a singular behavior at the
monopole's position, $r=0$. In a more realistic model for the
global monopole, we should not expect this kind of divergences.
So, partly motivated by this idea, we decided to return to this
analysis considering a non-vanishing radius to the monopole core.
In addition, this investigation enables us to clarify the role of
the finite core effects on the induced self-interaction. The
latter is a consequence of the distortion on the particle's
electric field caused by the spacetime curvature and topology. For
the general spherically symmetric static model of the core with
finite thickness we have constructed the corresponding three
dimensional Green function in both exterior and interior regions.
In the region outside the core this function is presented as a sum
of two distinct contributions. The first one corresponds to the
Green function for the geometry of a point-like global monopole,
previously investigated in \cite{Mello3}, and the second one is
induced by the non-trivial structure of the monopole core. The
latter is given by formula (\ref{Gc}) with the coefficient from
(\ref{Djl}). This coefficient is determined by the interior radial
solution regular at the core center and describes the influence of
the core properties on the physical characteristics in the
exterior region. The electrostatic self-energy and self-force are
obtained from the Green function in the coincidence limit after
the subtraction of the corresponding divergent part. This
procedure corresponds to the renormalization of the particle mass.
For points very far away from the core the most relevant
contribution is given by the first part of the self-energy, while
for points near the core surface the most relevant part is
represented by the second contribution. For the particle inside
the monopole core the electrostatic self energy is given by
formula (\ref{Urenin}), where the first term on the right is the
self energy for the background geometry described by the line
element (\ref{gm2}) for all values $r_{c}\leqslant r<\infty $, and
the second term is due to the global monopole geometry in the
region~$r>a$.

As an example of the application of the general results, in section \ref%
{sec:Flower} we have considered a simple core model with a flat spacetime
inside the core, so called flower-pot model. In this model, the
self-energies in the exterior and interior regions are given by formulae (%
\ref{UrenFlow}) and (\ref{Gr-in1}), respectively. The corresponding
self-forces are repulsive with respect to the core boundary in the case $%
\alpha <1$ and attractive for $\alpha >1$. In particular, for the first
case, the charge placed at the core center is in a stable equilibrium
position. Similar results can be obtained for the case of a point-like
magnetic charge. We have investigated the expressions for the self energies
in various asymptotic regions of the parameters. In particular, it has been
shown that for large values $\alpha $ the renormalized self-energy in the
exterior region tends to a finite limiting value. For small values $\alpha $%
, the core-induced part is exponentially suppressed by the factor $\exp [2%
\sqrt{2}\ln (a/r_{0})/\alpha ]/\alpha $, while the point-like
monopole part behaves like $1/\alpha ^{2}$ and, hence, the latter
strongly dominates. Although in the flower-pot model, we have
found a finite value of the self-energy at the monopole's center,
it presents a logarithmical singular behavior at the core
boundary. In this way the singular behavior becomes softer than
for the point-like monopole case. The corresponding divergences
are related to the idealization assuming that the transition range
between the interior and exterior geometries has zero thickness.
We expect that the results obtained in the present  paper will be
also valid in a more realistic model at distances from the
transition range much greater than the thickness of this range.

\section*{Acknowledgment}

AAS was supported by PVE/CAPES Program and in part by the Armenian Ministry
of Education and Science Grant No. 0124. ERBM thanks Conselho Nacional de
Desenvolvimento Cient\'\i fico e Tecnol\'ogico (CNPq) and FAPESQ-PB/CNPq
(PRONEX) for partial financial support.

\end{document}